# Study on FPGA SEU Mitigation for Readout Electronics of DAMPE BGO Calorimeter

Zhongtao Shen, Changqing Feng, Shanshan Gao, Deliang Zhang, Di Jiang, Shubin Liu, Qi An *Abstract*–The BGO calorimeter, which provides a wide measurement range of the primary cosmic ray spectrum, is a key sub-detector of Dark Matter Particle Explorer (DAMPE). The readout electronics of calorimeter consists of 16 pieces of Actel ProASIC Plus FLASH-based FPGA, of which the design-level flip-flops and embedded block RAMs are single event upset (SEU) sensitive in the harsh space environment. Therefore to comply with radiation hardness assurance (RHA), SEU mitigation methods, including partial triple modular redundancy (TMR), CRC checksum, and multi-domain reset are analyzed and tested by the heavy-ion beam test. Composed of multi-level redundancy, a FPGA design with the characteristics of SEU tolerance and low resource consumption is implemented for the readout electronics.

## I. INTRODUCTION

### A. The Dark Matter Particle Explorer

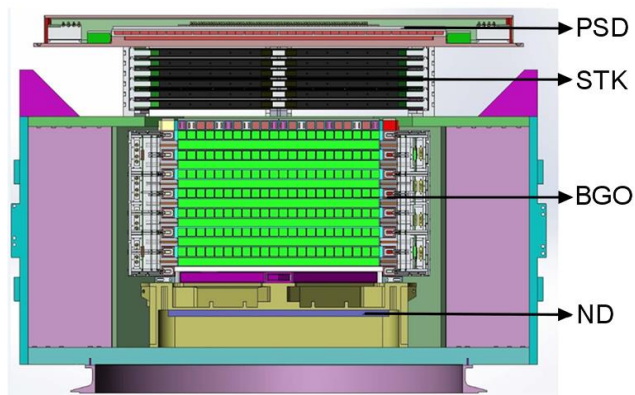

Fig. 1 DAMPE Detector Cross Section

The Dark Matter Particle Explorer (DAMPE) is being constructed as a scientific satellite to search for the proof of the existence of Dark Matter in space. As shown in Fig. 1, the DAMPE consists of four sub-detectors: a plastic scintillator detector (PSD), a silicon tracker (STK), a BGO calorimeter (BGO) and a neutron detector (ND) [1], [2]. Besides, there is a trigger board providing trigger signal for the four sub-detectors and a controlling computer in charge of controlling and corresponding with the four sub-detectors. As the satellite is designed to fly on a near-earth orbit with the altitude of 500km for more than 3 years, the radiation damage effects of semiconductor caused by high energy particles in space environment is one main threat to the reliability of the space electronics.

The BGO calorimeter is in charge of observation of high-energy electrons/positrons and gamma rays [3]. The readout electronics system of BGO consists of 16 front end electronics (FEE) boards. And on each of the FEE board, a FPGA works as the controlling chip, in charge of handling the command sent to the FEE, controlling the work status of other chips on the FEE board, acquiring, packaging and sending the scientific data and engineering parameter. A flash-based FPGA of Actel, ProASIC Plus (APA) is chosen as the BGO FEE controlling chip.

### B. PROASIC PLUS (APA)

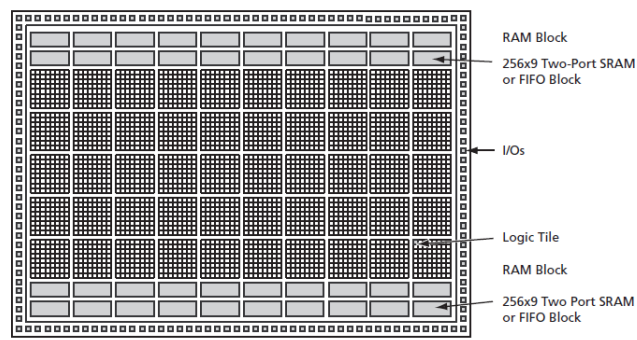

Fig. 2 Structure of APA FGPA

The APA family which adopts 0.22um LVCMOS process with four layers of metal has abundant programmable resource such as logic tiles, global nets, embedded RAMs and I/Os, as shown in Fig. 2. And it uses a live-at-power-up ISP flash switch as its programming element [4].

In the switch, two transistors share the floating gate, which stores the programming information and the upset mechanism for a heavy ion is to discharge the floating gate by generating charge in the bottom and top oxides that diffuse to the floating gate. However, the amount of charge generated by an ion with linear energy transfer (LET) value of 37 MeV $cm^2$/mg is less than 1% of the total charge on a programmed floating gate [5]. Therefore the configuration unit is insensitive to SEE and the programming information storing in logic tiles and I/Os is unlikely to be changed when the chip works in space.

Manuscript received June 16, 2014. This work was supported by the Strategic Priority Research Program on Space Science of the Chinese Academy of Sciences (Grant No. XDA04040202-4), and the National Basic Research Program (973 Program) of China (Grant No. 2010CB833002).

All Authors are with State Key Laboratory of Particle Detection and Electronics, University of Science and Technology of China, No.96, Jinzhai Road, Hefei, Anhui, China. (telephone: +86-0551-63600408, e-mail: henzt@mail.ustc.edu.cn,fengcq@ustc.edu.cn,shawn@ustc.edu.cn,dlzhang@mail.ustc.edu.cn,jiangdi@mail.ustc.edu.cn,liushb@ustc.edu.cn,anqi@ustc.edu.cn ).

However, experiments also show that the D Type Flip-Flop (DFF) configured from the logic tile and the embedded RAMs are sensitive to SEU. Test result shows that the SEU LET value for DFF is less than 3 MeV cm$^2$/mg and using this value CREME96 predicts that the SEU probability for the chip is about $6.8 \times 10^{-7}$ bit$^{-1}$ day$^{-1}$ at the altitude of 500km [5], [6]. Considering that there are about 2500 memory units in the APA chip and one-bit error can cause software error or even infinite loop status, it is necessary to take some measures to mitigate SEU when ProASIC Plus device is used in space.

## II. SEU Mitigation Techniques

### A. Triple Modular Redundancy

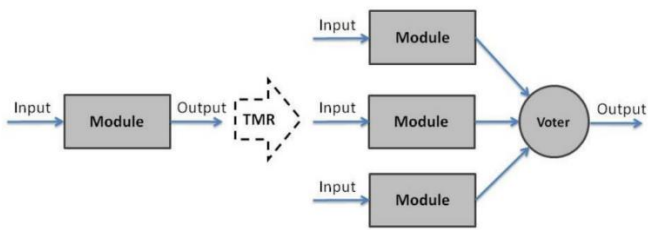

Fig. 3 Structure of TMR

The Triple Modular Redundancy (TMR) technique is the most commonly adopted passive hardware redundancy technique achieving fault masking properties [7]. It uses three replicas for one memory cell and adds a voter that identifies the correct result among the three ones on the basis of a majority vote, as shown in Fig. 3. The technique may be applied at different levels, from the whole system to a single register.

The TMR technique improve the reliability of system, but it cause some problems at the same time such as much more resource consumption, speed reduction, power increasing, hardness in placing and routing and so on. To achieve maximum SEU tolerance with appropriate engineering consumption, usually TMR is only used in key registers and RAMs.

### B. Error Detection and Correction

Error detection and correction codes (EDAC) are often used to improve the reliability of data storage media. The general idea for achieving error detection and correction is to add some redundancy, and the redundancy is a fixed number of check bits, which are derived from the data bits by some deterministic algorithm. The error detection schemes include Parity bits, Hamming Code, Cyclic redundancy check and so on.

## III. SEU Mitigation Applied in FEE FPGA Design

### A. Structure of FPGA of BGO FEE

The FPGA of BGO FEE mainly consists of four parts: the scientific data acquisition part, the control part, the monitor part and the status manager part, as shown in Fig. 4. The scientific data acquisition part is in charge of communicating with the peripheral chips which are related to data acquisition, caching the data got from the peripheral devices, packaging the scientific data and sending the data package. The control part is in charge of receiving commands sent to FPGA, judging the validity of the command, executing the command and giving the response. These two parts which achieve the main function of FPGA take up about 78% resource of all logic consumption.

For the control part, there is a command handling procedure which starts when receiving command and completes command analysis, executing and responding step by step. And for the scientific data acquisition part, there is a scientific data acquisition procedure which is triggered by the trigger signal from the trigger board and also consists of many steps. When a procedure finishes, the part of the logic is idle until next command or trigger signal comes.

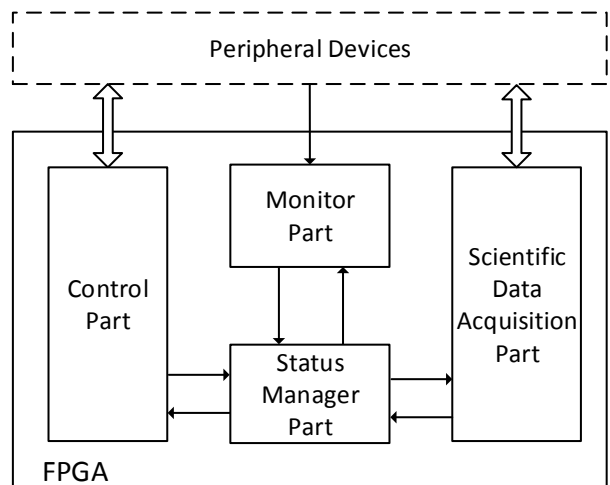

Fig. 4 Structure of FPGA of BGO FEE

### B. Multi-Domain Reset and Multi-Level Reset

As mentioned before, the logic has a command handling procedure and a scientific data acquisition procedure. In space, SEU can happen in any step of the procedure. And this may lead the control part or the scientific data acquisition part into an infinite loop status, at which the system can't work normally and need to be reset. However, it is not necessary to reset the whole system if only parts of the logic are under infinite loop condition. Therefore, Multi-domain reset and Multi-level reset is adopted in FPGA design.

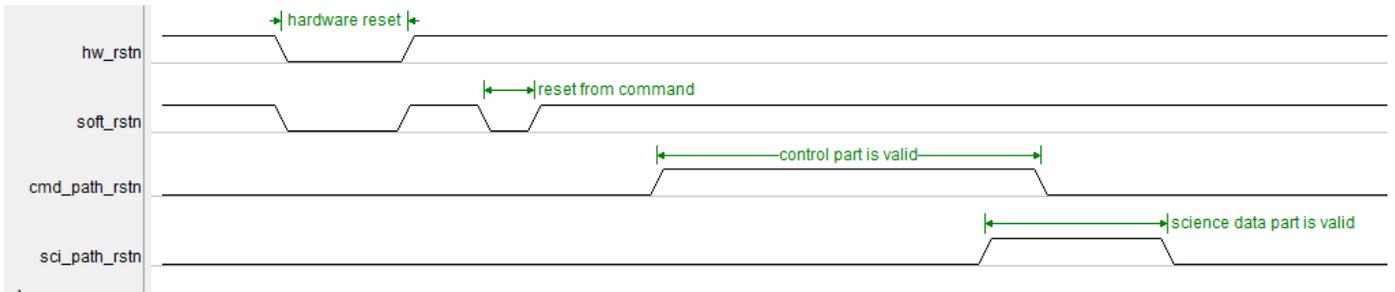

Fig. 5 Multi-domain reset signal

As shown in Fig. 5, four reset signals are used in logic design including hw_rstn, soft_rstn, cmd_path_rstn and sci_path_rstn. Hw_rstn is the hardware reset signal which comes from the outside reset chip. It resets all registers and RAMs to their default and is used to initialize the FPGA when powering on. Soft_rstn comes from reset command and resets all registers and RAMs expect a shifter which is used in command reset. Cmd_path_rstn and sci_path_rstn is used to reset the control part and the scientific data acquisition part respectively. When a data acquisition process begins, a timer starts to count. If the data acquisition process doesn't finish in a certain time because of SEU or something else, the sci_path_rstn actives and resets the scientific data acquisition part automatically. This is the principle of sci_path_rstn, and the principle of cmd_path_rstn is similar. These two reset signals prevent the system entering infinite loop status and guarantee that the SEU happening in the previous process doesn't influence the next one. Besides, when the scientific data acquisition part and the control part are idle, the reset signals are active and set the two parts into sleep, which can effectively avoid the influence of SEU.

*C. TMR and CRC*

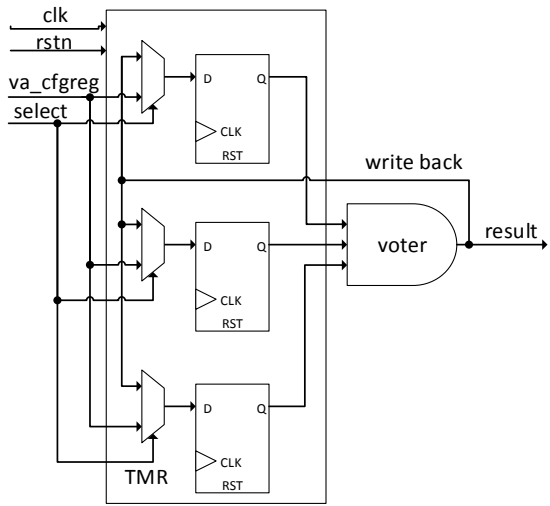

Fig. 6 TMR structure of va_cfgreg

As mentioned before, most of the registers in the logic belongs to the control part or the scientific data acquisition part and they are reset when the procedure is done. However, there're some key registers and RAMs which are valid all the time and only respond to the global reset signal. Therefore TMR technique is used in vital registers and RAMs to make sure that the values in them are correct.

Fig. 6 shows the TMR structure of va_cfgreg, which is one of the key registers in FPGA. In the structure, each replica is a DFF and the voter consists of combinational logic circuits. The input of each DFF is connected with a 2:1 multiplexer, which decides whether the outside signal or the voter result is the input of the DFF. When the register doesn't need to be rewritten, the voter result is chosen as the inputs of the three DFFs and the values in them are refreshed by the voter result at each rising edge of clock signal. Therefore, if SEU happened in one DFF, the voter result which is decided by the majority of the three DFFs is still correct and the wrong value would be refreshed by the correct value immediately.

To avoid that Mutiple Bit Upset (MBU) happens in two or three replicas of one TMR unit, before automatic placement and routing, three replicas are manually set in different physical areas.

To make further efforts to ensure the correctness of the value in key registers, the values are monitored as engineering parameter all the time. However, unlike the ones in registers, the data in RAMs are so many that they can't be monitored in real-time. Therefore, the cyclic redundancy check (CRC) is adopted to automatically detect data corruption in RAMs.

The CRC result is calculated before and attached at the end of the RAM. When the data in RAM is used, the CRC result is calculated again and compared with the one at the end of the RAM. The mismatch of the two CRC results shows the data corruption in RAM and an indicating bit is active automatically. The indicating bit is also monitored as engineering parameter and is used as a call for RAM reconfiguration.

IV. TEST AND RESULTS

After using these means in the logic design, an ion-beam testing is performed at the Heavy Ion Research Facility in Lanzhou (HIRFL) cyclotrons to evaluate to the SEU tolerance of FPGA logic, as shown in Fig. 7.

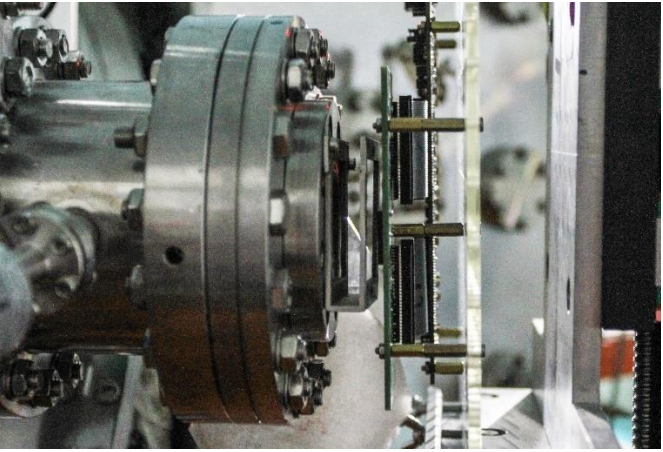
Fig. 7 SEU Test Site

## A. Test Platform

The test system consists of a host PC, a Master Board and a DUT Board. A LabVIEW program works on the host PC to control the whole system and monitor the status of APA. The Master board which works in the environment without radiation is in charge of communicating with host computer and the DUT Board and powering up the DUT Board. The APA chip is placed on the DUT Board and only the DUT Board is placed under radiation condition to make sure that the system function well while APA is under radiation test.

## B. Irradiation Test with High LET Value

The test is performed at HIRFL-TR5 terminal, using Bismuth ions. Irradiations are conducted in air, at ambient temperature, with heavy ions passing through a vacuum/air transition foil. By the means of changing the thickness of air, the LET value is adjusted to about 90 Mev cm$^2$/mg which is very high to chips used in space. An APA600 chip with removal of the package lid is configured with a simple logic and tested under the radiation condition.

APA600 works under the irradiation for about 10 minutes and the total fluence reaches up to $8.4 \times 10^6$ ions/cm$^2$. During this time, the supply current of FPGA is monitored and no abnormal current is found, which means no SEL happens in the experiment and the APA chips are insensitive to SEL. Besides, the FPGA works well during the experiment and no configuration damage is observed, which means the structure of flash switch in APA is immune to SEU and the configuration information in APA won't be change when it works in space. The experiment results correspond with the research made by Gregory R. Allen and Gary M. Swift [5].

## C. Functional Test

Configured with a lite version of the whole system logic, APA300 is put on the condition with the LET value of 39.6 MeV*cm$^2$/mg and the ion flux of 100 ions/cm$^2$/s. The experiment continues for about 25 minutes and the FPGA works with no mistakes. When receiving commands, the control part can handle it without error and the science data acquisition part also works well.

Through the experiment, the three ways used in the logic design for SEU mitigation are proved to be effective and the mitigation techniques make sure that the APA chip can work in the space environment.

## V. DISCUSSION

### A. The Reliability of TMR with Correction

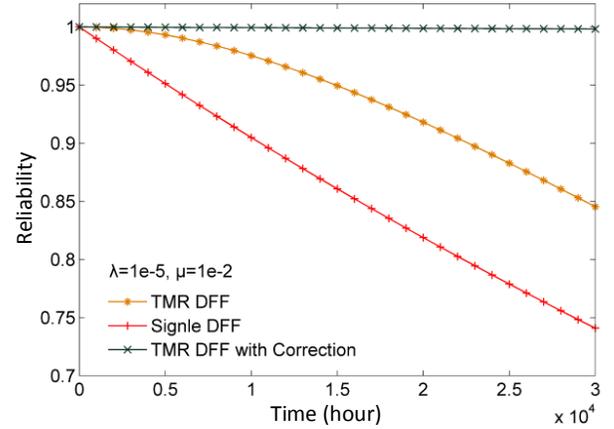
Fig. 8 Reliability Change with Time

Using Markov model in fault-tolerant computing, a system's reliability can be predict [8]. Fig. 8 shows reliability change of three structures with time. Compared with the one of TMR DFF without correction and signal DFF, the reliability of TMR DFF almost doesn't change with time and still very close to 1 at 30000 hours later. Therefore, the structure of TMR with correction can highly improve the reliability of a register.

### B. The Reliability of the Ion Test

According to the test results, the SEU LET for APA chip is about 3 MeV cm$^2$/mg [5], and at the attitude of 500km, the flux at this LET value is less than 0.5 ions/cm$^2$/s [6]. Compared with the flux of 100 ions/cm$^2$/s in test environment at HIRFL-TR5, the flux in the space environment is very low. Besides, as discussed above, due to the write-back line in the TMR structure, the SEU error won't accumulate by time. Therefore, the ion test simulates a much harsh condition and verifies that the FPGA logic with SEU mitigation techniques has the ability of SEU tolerance in the space environment.

### C. Resource Consumption and Performance Reduction

The methods of SEU mitigation have efficiently declined the sensitivity to SEU of FPGA. However, as mentioned before, resource consumption and speed degradation also need to be considered when logic hardening techniques are used.

Table I shows comparison of the logic tile consumption, highest work frequency and power between the logic with hardening techniques and the one without hardening techniques. From the table we can see that due to the SEU mitigation design, the logic tile increases from 57.1% to 77.0% which is still very low, and the highest work frequency declines to 33.2 MHz which also meets the requirements of design. The power, which increases from 124.1 mW to 153.8 mW, will not cause a problem either. Therefore, not only the techniques used in the logic design mitigate SEU effectively but also the resources they consume is acceptable.

TABLE I
COMPARISON OF LOGIC WITH AND WITHOUT HARDENING

|  | Logic Tile/% | Highest Frequency/MHz | Power/mW |
|---|---|---|---|
| With Hardening Techniques | 57.1 | 44.1 | 124.1 |
| Without Hardening Techniques | 77.0 | 33.2 | 153.8 |

## VI. CONCLUSION

In order to enable the APA to function well in the space radiation environment for DAMPE, SEU mitigation techniques are used in the logic. The SEU mitigation means include TMR, CRC and Multi-domain reset. After logic hardening, an ion-beam experiment is carried out and it proves that the SEU mitigation techniques are effective and the FPGA after logic reinforcement works under the condition with radiation with no mistakes.

REFERENCES

[1] Chang, J., et al. "An excess of cosmic ray electrons at energies of 300–800 GeV," Nature 456.7220 (2008): 362-365.
[2] J. Chang, "Dark Matter Particles Detection in Space," Journal of Engineering Studies. vol.2(2), pp. 95-99, June 2010.
[3] Zhang Y L, Li B, Feng C Q, et al. A high dynamic range readout unit for a calorimeter[J]. Chinese Physics C (HEP & NP), 2012, 36(1): 71-73.
[4] ProASIC A, FPGA E P. Features and Advantages http://www. actel. com/documents[J]. PA3_E_Tech_WP. pdf.
[5] Allen G R, Swift G M. Single event effects test results for advanced field programmable gate arrays[C]//Radiation Effects Data Workshop, 2006 IEEE. IEEE, 2006: 115-120.
[6] Tylka A J, Adams J H, Boberg P R, et al. CREME96: A revision of the cosmic ray effects on micro-electronics code[J]. Nuclear Science, IEEE Transactions on, 1997, 44(6): 2150-2160.
[7] Siewiorek D P, Swarz R S. The theory and practice of reliable system design[M]. Bedford, MA: Digital press, 1982.
[8] McMurtrey D, Morgan K, Pratt B, et al. Estimating TMR reliability on FPGAs using markov models[J]. BYU Dept. Electr. Comput. Eng., Tech. Rep, 2006.